\documentstyle[11pt]{article}
\begin{document}
\renewcommand{\thesection}{\Roman{section}.}
\renewcommand{\theequation}{\arabic{section}.\arabic{equation}}
\parskip = 0 pt
\baselineskip = 23 pt
\parindent = 15 pt
\abovedisplayskip=13pt plus 3pt minus 9pt
\belowdisplayskip=13pt plus 3pt minus 9pt
\vskip 15  pt
\centerline{\LARGE Dynamical Semigroup Description of Coherent}\par
\centerline{\LARGE and Incoherent Particle-Matter Interaction}\par
\vskip 15  pt
\centerline{L.~Lanz\footnote{Dipartimento di Fisica
dell'Universit\`a di Milano and Istituto Nazionale di Fisica
Nucleare, Sezione di Milano, Via Celoria 16, I-20133, Milan,
Italy. E-mail: lanz@mi.infn.it} and B.~Vacchini\footnote{Dipartimento di 
Fisica
dell'Universit\`a di Milano and Istituto Nazionale di Fisica
Nucleare, Sezione di Milano, Via Celoria 16, I-20133, Milan,
Italy. E-mail: vacchini@mi.infn.it}}
\vskip 15 pt
\centerline{\sc Abstract}\par
\vskip 15 pt
{
\baselineskip=12pt
The meaning of statistical experiments with single microsystems
in quantum mechanics is discussed and a general model in the
framework of non-relativistic quantum field theory is proposed,
to describe both coherent and incoherent interaction of a single
microsystem with matter.
Compactly developing the calculations with
superoperators, it is shown that
the introduction of a time scale,
linked
to irreversibility of the reduced dynamics,
directly leads
to a dynamical semigroup expressed in terms of quantities typical
of scattering theory.
Its generator consists of two terms, the first linked to
a coherent wavelike behaviour, the second related to an
interaction having a measuring character, possibly connected to
events the microsystem produces propagating inside matter.
In case these events breed a measurement, an explicit
realization of some
concepts of modern quantum 
mechanics (``effects'' and ``operations'') arises.
The relevance of this description to
a recent debate questioning the validity
of ordinary quantum mechanics to account
for such experimental situations as, e.g., neutron-interferometry,
is briefly discussed.
}\par
\vskip 25 pt
\hrule
\par
\vfill\eject
\par
\section{INTRODUCTION}
\par
\vskip 10pt
Consider a source, emitting practically only one particle each
time, feeding an interferometer; one of the most impressing
features of quantum mechanics is the fact that  the record in a
detector of the output of the interferometer, during a suitable
time interval, shows an interference pattern. If the experimental
set-up allows detectable events to be produced during the time the
particle takes to pass through the interferometer, thus showing
which way the particle went, a two component pattern is found,
respectively affected and not affected by interference. Seemingly
the interfering part can be strongly attenuated, if the
probability of detecting events is enhanced, still retaining its
visibility. Let us mention some of the experiments of relevance
to the question carried out in different fields in the last
years
(Rauch, 1990, 1995; Rauch {\it et al.}, 1990; Mittelstaedt {\it et
al.}, 1987; Chapman {\it et al.}, 1995).
It was sometimes claimed and also written in textbooks, that the
very possibility of such a detection  forces the interference
pattern to disappear; such somewhat strange expectation is rooted
in an exaggerated faith in the so called state reduction
postulate of quantum mechanics. This postulate is a strongly
idealized description of what happens to a quantum system due to
the interaction with a device measuring a given observable of
the system; using this postulate a shorthand explanation of
measurement is usually given, based on the idea that a quantum
system must be represented by a ``state vector'' $\psi(t)$.
A much more comfortable situation is met if, instead of a state
vector, a statistical operator ${\varrho} (t)$
is taken
as the
basic mathematical representation of a quantum system (Lanz,1994).
This attitude is sometimes considered suitable for applications,
e.g. quantum optics, but not fine enough for more
fundamental problems; it is often implicitly assumed that a
statistical operator applies only to the description of a
statistical mixture of a large number of microsystems, while in
modern experiments often only one or very few relevant
microsystems are present altogether in the experimental device.
In this single-particle experiments it is often argued
(Namiki and Pascazio, 1991; Thomson, 1993)
that the system is to be described by a state vector.
In our opinion, instead, one-particle quantum
mechanics,
no matter if one uses $\psi(t)$ or ${\varrho}(t)$,
refers in principle to  a statistical experiment in which
repeatedly a single particle is produced, prepared and observed
under fixed macroscopic conditions; this does not oppose the fact
that a beam of particles whose interactions are negligible and
whose correlations are irrelevant may be treated in many
experimental situations as effectively equivalent to the former
preparation. It is just the modalities of the statistical
experiment, which remain unchanged during the different runs of
the experiment, that are represented by the statistical operator
(or by the state vector, when this higher idealization works);
this is indeed the striking difference with classical mechanics,
where to each run of the statistical experiment corresponds a
trajectory in phase space. In this context a completely different
point of view seems to underlie the so-called 
many-Hilbert-space quantum mechanics, that was recently
proposed
(Namiki and Pascazio, 1993).
In this framework a wave function is
associated to each single-run of a statistical experiment
and for example in a Young's interference experiment random phase
shifts between the two branch waves may arise in the repeated
experimental runs, due to interaction with matter along one of
the two branches, leading to attenuation of the interference
pattern
(Namiki and Pascazio, 1991).
\par
As it is well known state vectors $\psi \in {\cal H}$, via the one
dimensional projections ${P}_{\psi}$ on ${\cal H}$, correspond to the
subset of extreme points of the convex set ${\cal K}$ of statistical
operators in ${\cal H}$: i.e. they cannot be interpreted as mixtures
of other possible preparations and any ${\varrho} \in {\cal K}$ can
be
represented as ${\varrho} = \sum_j p_j {P}_{\psi_j}$. For this reason
state vectors $\psi \in {\cal H}$ are also called ``pure states''.
Let us
recall a relevant
mathematical result (Davies, 1976);
any
invertible affine mapping ${\cal M}$ on
${\cal K}$
onto ${\cal K}$ has the form:
        \[
        {\cal M} {\varrho} =
        {M} {\varrho} {M}^{{\scriptscriptstyle \dagger}},
        \]
${M}$ being a unitary (or antiunitary)
operator on ${\cal H}$; then, if time evolution
is represented by such a mapping
(Comi {\it et al.}, 1975),
the
basic role of pure
states for the dynamics becomes obvious and consequently also the
relevance of the Schr\"odinger equation, of the Hamilton operator
and finally the correspondence with classical mechanics and
classical field theory. Summing up in
formulae:
        \[
        {\varrho}_t = {\cal M}_{tt_0}
        {\varrho}_{t_0}={U}(t,t_0)
        {\varrho}_{t_0}{U}^{{\scriptscriptstyle \dagger}}(t,t_0)=
        \sum_j p_j {P}_{\psi_j(t)}
        \]
        \[
        \psi_t = {U}(t,t_0) \psi_{t_0} , \qquad  i\hbar
        {
        d\psi_t
        \over
              dt
        }       = {H}_t \psi_t .
        \]
In fact the main part of the physics of 
microsystems can
be developed almost neglecting the concept of statistical
operator (a noteworthy exception however is given by the
definition of the quantum collision cross-section, Taylor, 1972;
Ludwig, 1976).
\par
Such a reversible dynamics is to be expected for an isolated
system.
If interaction with an
environment is not negligible during the time evolution the
question is to be raised if this evolution can be simply
described by a mapping ${\cal M}_{tt_0}$ on ${\cal K}$; i.e. if
${\varrho}_t$ is
uniquely determined by ${\varrho}_{t_0}$ and not by the whole history
$\left \{ {{\varrho}_{t'}; t' \leq t_0} \right \}$ before $t_0$,
recorded via interaction by this environment.
In this general situation the system becomes the whole complex of
particle plus environment and no disentanglement of the
particle's degrees of freedom
is possible. On the contrary a neat and extremely relevant
simplification occurs if such a mapping ${\cal M}_{tt_0}$ exists:
then the one-particle Hilbert space ${\cal H}$  and not the
Fock space of the whole
system is the relevant
mathematical framework.
Let us assume that
this simplification occurs, typically due to the fact
that the aforementioned history is forgotten during the
time elapsed before ${\varrho}_t$ varies appreciably, as in the
case of markovian dynamics;
nevertheless one can
no longer expect ${\cal M}_{tt_0}$ to be invertible: then the
statistical operator ${\varrho}_t$ acquires a primary role. In
differential form the evolution equation for ${\varrho}_t$ is:
        \[
        {
        d{\varrho}_t
        \over
              dt
        }       = {\cal L}_t {\varrho}_t  , \qquad {\cal L}_t
        = \lim_{\tau \to 0} {
        {\cal M}(t+\tau , t) - {\cal I}
        \over
                                  \tau
        }                             ,
        \]
        \begin{equation}
        {\cal M}_{tt_0} = {\rm T}
        \left(
        {\exp \int_{t_0}^{t} \, d
        t'  \, {\cal L}(t')}
        \right)
         .  \label{i3}
        \end{equation}
In $\S\,2$
we explicitly construct the generator ${\cal L}_t$ of the
temporal evolution for the microsystem showing in a general
way how it can be obtained starting from the
Hamiltonian describing the local interaction between microsystem
and macrosystem. An essential step is the introduction of a time
scale on which the system is to be described, linked to the
irreversibility of the interaction.
To develop the calculations we rely upon a reformulation of the
theory of scattering based on superoperators, that is mappings
defined on the algebra generated by creation and destruction
operators acting in the Fock space. Quantum statistics is readily
accounted for and the mapping ${\cal T}(z)$ [see (\ref{211})],
strictly connected to the
transition operator of the quantum theory of scattering, plays a
central role from the very beginning.
The use of the Heisenberg picture, consistent with the
concentration of one's attention on the microsystem's
observables, allows to keep the whole complex structure of the
macrosystem into account.
The generator obtained is of the Lindblad type, though allowing
for unbounded operators.
The general structure of such generators,  ensuring
that ${\cal M}_{tt_0}$ maps
${\cal K}$
into ${\cal K}$,
is the following:
        \begin{equation}
        {\cal L}_t{\varrho} = -{i \over \hbar}
        \left(
        {{{H}}_t {\varrho} -{\varrho} {{H}}_t}
        \right)
         -
        { {1\over\hbar}}
        \left(
        {{A}_t {\varrho} + {\varrho} {A}_t}\right)
        +{1 \over \hbar} \sum_j
        {{L}}_{tj}{\varrho}
        {{L}}_{tj}^{{\scriptscriptstyle \dagger}}  \label{i4}
        \end{equation}
        \[
        {{H}}_t ={{H}}_t^{\scriptscriptstyle \dagger} , \quad {A}_t
        \geq 0, \quad
        {L}_{tj}
        {{\rm \hbox{\quad{being operators in ${\cal H}$}  \quad}}}
        .
        \]
The relation:
        \begin{equation}
        {A}_t =  {1\over2}     \sum_j
                 {{L}}_{tj}^{{\scriptscriptstyle \dagger}}{{L}}_{tj},
                 \label{i5}
        \end{equation}
must be satisfied in order that ${\hbox{\rm Tr}}
{\varrho}_t$
be conserved. If
the particle can be absorbed
(\ref{i5}) is replaced
by
        \begin{equation}
        {A}_t \geq {1\over2}     \sum_j
        {{L}}_{tj}^{{\scriptscriptstyle \dagger}}{{L}}_{tj}
        ,  . \label{i5primo}
        \end{equation}
If the last term in  (\ref{i4}) is neglected, for a pure state
${\varrho}_t = {\vert \psi_t \rangle} {\langle \psi_t \vert}$
(\ref{i3}) yields the
Schr\"odinger equation:
        \begin{equation}
        i\hbar
        {
        d\psi_t
        \over
              dt
        }       =
        \left(
        {{{H}}_t -i {A}_t}
        \right)
              \psi_t;
        \label{i6}
        \end{equation}
this is the basis for the wavelike description of propagation of a
particle inside matter.
Setting ${{H}}_t -i {A}_t = { \textstyle{{{\bf p}}^2} \over \textstyle
2m}+V({\bf x},t)$ one can define
        \begin{equation}
        n({\bf x},\nu,t)= \sqrt{1- {
        V({\bf x},t)
        \over
        h\nu        }}                             \label{i6bis}
        \end{equation}
as refractive index of the medium, where $h\nu$ is to be identified with the
energy of the incoming particle:
such a description is usually adopted in interferometric
experiments to explain how a block of matter, whose properties
are accounted for by the phenomenological macroscopic potential
$V({\bf x},t)$, placed in one of the two branches can induce a phase
shift in the corresponding branch-wave, or, in the case of an
imaginary potential, cause absorption.
Only in the very special case
of ${A}_t=0$, i.e. for a real ``macroscopic'' potential $V({\bf x},t)$,
by  (\ref{i5}) or (\ref{i5primo}) one has ${L}_{tj}=0$ and
(\ref{i6}) is
exactly equivalent to (\ref{i4}). In presence of absorption ${A}_t \neq
0$ implies by  (\ref{i5}) ${L}_{tj} \neq 0$ for some $j$; but also in
absence of absorption one cannot expect that ${L}_{tj}=0$. Notice
that, if one is not aware of the basic role of  (\ref{i4}) and of the
importance of the last term at its r.h.s., by  (\ref{i6}) one could
be confirmed in the erroneous belief that non-reality of the
potential $V$ is exclusively linked to absorption processes.
To grasp the significance of the term ${1 \over \hbar} \sum_j
        {{L}}_{tj}{\varrho}
        {{L}}_{tj}^{{\scriptscriptstyle \dagger}}$
        for the dynamics of ${\varrho}$ let us
write the evolution of ${\varrho}$ due to it in a small time
interval $\tau$ in the form:
        \begin{equation}
        \label{misura}
        \Delta {\varrho}= {
        \tau
        \over
            \hbar
        }        {\hbox{\rm Tr}}
        \left(
        {2 {A}_t {\varrho}}
        \right)
        \sum_j
        {\tilde  L}_{tj}{\varrho}
        {\tilde  L}_{tj}^{{\scriptscriptstyle \dagger}}, \qquad
        {\tilde  L}_{tj}=    	{
        {{L}}_{tj}
        \over
        \sqrt{{\hbox{\rm Tr}}
        \left(
        {2 {A}_t {\varrho}}
        \right)
        }
        }       ;
        \end{equation}
The statistical operator $\sum_j {\tilde L}_{tj}{\varrho}
        {\tilde  L}_{tj}^{{\scriptscriptstyle \dagger}}$
is a  mixture of
subcollections ${\tilde  L}_{tj}{\varrho}
        {\tilde  L}_{tj}^{{\scriptscriptstyle \dagger}}$
related to outcome channels labeled by the index $j$; it
bears some resemblance with the statistical operator $\sum_j
{P}_j {\varrho} {P}_j$ which represents, by the previously mentioned
reduction postulate, the system after the measurement of an
observable ${A} = \sum_j a_j {P}_j$;
${ 1  \over
            \hbar
        }        {\hbox{\rm Tr}}
        \left(
        {2 {A}_t {\varrho}}
        \right)
        $
expresses the strength of the coupling to the incoherent regime.
More generally a mapping whose infinitesimal generator is of the
form (\ref{i4}) admits measuring decompositions that have been
characterized in the context of ``continuous measurement
theory'', initiated  by   Davies for the counting processes and
developed later in full generality (for a recent review see
Lanz and Melsheimer, 1993   and Lanz, 1994).
These decompositions are related to the operators ${L}_{tj}$,
responsible for the irreversible dynamics, and clarify what is
meant by the measuring character of a mapping describing the
temporal evolution of a system.
We will see in $\S\,3$ that  (\ref{i4}) couples very simply the
typical wave dynamics, which is responsible for  interference
phenomena, with a ``non-coherent'' regime.
Obviously in many instances the main interest is to put
the wavelike behaviour in major evidence; this amounts to make
${L}_{tj}$ negligible, so that  (\ref{i6}) is indeed suitable to
describe the dynamics. On the contrary more recent
investigations, e.g. neutron interferometry in presence of stray
absorption in one path of the
interferometer
(Rauch 1990, 1995; Rauch {\it et al.}, 1990),
 aim at
investigating the competition
between wavelike coherent behaviour and which-way detection:
then  (\ref{i3}) and (\ref{i4}) must be considered.
In $\S\,3$ the physical interpretation
of the dynamics thus obtained for the microsystem is discussed,
showing the interplay between a ``purely optical'' regime [such
as in  (\ref{i6}) and (\ref{i6bis})] and an ``events producing'' one,
strictly connected to the presence of the incoherent contribution
in the r.h.s. of (\ref{i4}).
\vskip 3 truecm
\par
\section{CONSTRUCTION OF THE GENERATOR}
\par
\setcounter{equation}{0}
\vskip 10pt
We assume for simplicity that the whole system is confined, e.g.,
in a box; eventually we can get rid of this confinement
letting the size of the box go to infinity.
The  microsystem is described in a Hilbert space ${{\cal H}^{(1)}}$;
energy eigenvalues are ${E_f}$, energy eigenstates
$u_f$, spanning the space ${{\cal H}^{(1)}}$.
In this paper we shall make use of the formalism of
non-relativistic quantum
field
theory, which will prove to
play an essential role in order to obtain a general procedure
leading from the
second quantized
Hamiltonian ${H}$ of the whole
system, acting in the global Fock space
${{{\cal H}_{\scriptscriptstyle F}}}$,
to the generator of the semigroup ${\cal L}$ acting in
${\cal T}({{\cal H}^{(1)}})$ (the set of trace-class
operators in ${{\cal H}^{(1)}})$.
\par
We shall set:
        \[
        {H}={H}_0 + {H}_{\rm m} + {V}
        \]
        \[
        {H}_0 = \sum_f
        {E_f} {a^{\scriptscriptstyle \dagger}_{f}} {a_{{f}}} \qquad
         \qquad
        \left[{{a_{{f}}},{a^{\scriptscriptstyle
        \dagger}_{g}}}\right]_{\mp}=\delta_{fg}
        \]
where ${a_{{f}}}$ is the destruction operator for the microsystem,
either a fermi or a bose particle, in
the state $u_f$; ${H}_{\rm m}$ is the Hamilton operator for the
macrosystem ($\left[{{H}_{\rm m},{a_{{f}}}}\right]=0 $), also
containing the potential determining the internal structure of
the  macrosystem;
 ${V}$~represents
the interaction between the two systems.
We shall assume in this paper that no absorption process of the
microsystem occurs: then ${N}  =  \sum_{{h} }
{a^{\scriptscriptstyle \dagger}_{h}}  {a_{h}}$
is a constant, $[{N},{H}]=[{N},{V}]=0$.
The present treatment is non-relativistic due to the role
played by particle number conservation.
\par
We assume for the statistical operator the following expression:
        \begin{equation}
        {\varrho}=
        \sum_{{g} {f}}{}
        {a^{\scriptscriptstyle \dagger}_{g}}
        {{\varrho}^{\rm m}} {a_{{f}}}
        {{\varrho}}^{(1)}_{gf}    ,
        \label{25}
        \end{equation}
where ${{\varrho}^{\rm m}}$ is a statistical operator
in the subspace ${{{\cal H}^0_{\scriptscriptstyle F}}}$
of ${{{\cal H}_{\scriptscriptstyle F}}}$  in which
${N}=0$,
representing the
macrosystem
and therefore:
        \[
        {a_{{f}}}{{\varrho}^{\rm m}}=0 \quad
        {{\varrho}^{\rm m}}
        {a^{\scriptscriptstyle \dagger}_{f}}=0 \quad
        \forall f
        ,
        \]
while
${\varrho}$ is a statistical operator in the subspace
${{{\cal H}^1_{\scriptscriptstyle F}}}$  of
${{{\cal H}_{\scriptscriptstyle F}}}$  in which ${N}=1$.
As far as the microsystem is concerned, the dynamics of the
macrosystem is not appreciably perturbed by the presence of the
microsystem itself, so we can assume that
        \[
        {
        d{{\varrho}^{\rm m}(t)}
        \over
         dt
        }  =
        - {i \over \hbar} [{H}_{\rm m},{{\varrho}^{\rm m}(t)}]  .
        \]
The coefficients
${{\varrho}}^{(1)}_{gf}$ build a positive, trace one  matrix,
which can be considered as the
representative of a statistical operator ${{\varrho}}^{(1)}$ in
${{\cal H}^{(1)}}$.
In fact, since we are
interested in the subdynamics of the microsystem
and thus in observables of the form:
        \begin{equation}
        {A}
        = \sum_{h,k}
        {a^{\scriptscriptstyle \dagger}_{h}}
        {{\mbox{\sf A}}}^{(1)}_{hk}
        {a_{k}}   ,
        \label{23}
        \end{equation}
where ${{\mbox{\sf A}}}^{(1)}_{hk}$ is the matrix element of the
corresponding operator acting in ${{\cal H}^{(1)}}$,
we will make use of the following reduction formula from
${{{\cal H}_{\scriptscriptstyle F}}}$
to ${{\cal H}^{(1)}}$  for
the expectation value of an observable ${A}$
of
the form (\ref{23}) in the state (\ref{25}):
        \[
        {\hbox{\rm Tr}}_{{\cal H}_{\scriptscriptstyle F}}
        \left(
        {{A}{\varrho}}
        \right)
         = \sum_{h,k} {{\mbox{\sf A}}}^{(1)}_{hk}
        {{\varrho}}^{(1)}_{kh}=
        {\hbox{\rm Tr}}_{{{\cal H}^{(1)}}}
        \left(
        {{{\mbox{\sf A}}}^{(1)} {{\varrho}}^{(1)}}
        \right)
        \]
Considering in  particular the operator 
${A}=        {a^{\scriptscriptstyle \dagger}_{f}} {a_{g}} $
we have:
        \[
        {\hbox{\rm Tr}}_{{\cal H}_{\scriptscriptstyle F}}
        \left(
        {{A}{\varrho}}
        \right)
         =
        {{\varrho}}^{(1)}_{gf}
        .
        \]
To individuate the generator of the semigroup we will consider the
evolution of the statistical operator on a time scale $\tau$ much
longer than the correlation time for the  macrosystem, thus
approximating
$        {
        d {\varrho}_{gf}^{(1)}(t)
        \over
        dt
       }
$ by:
        \begin{equation}
        {
        \Delta {\varrho}_{gf}^{(1)}(t)
        \over
        \tau
        }
        =
        {1\over \tau}
        \left[
        {\varrho}_{gf}^{(1)}(t+\tau) -
        {\varrho}_{gf}^{(1)}(t)
        \right]
        =
        {1\over \tau}
        \left[
        {\hbox{\rm Tr}}_{{\cal H}_{\scriptscriptstyle F}}
        \left(
        {a^{\scriptscriptstyle \dagger}_{f}} {a_{g}}
        e^{-{{
        i
        \over
         \hbar
        }}H\tau}
        \varrho (t)
        e^{{{
        i
        \over
         \hbar
        }}H\tau}
        \right)
        -
        {{\varrho}}^{(1)}_{gf}(t)
        \right]
        .
        \label{new}
        \end{equation}
Exploiting the cyclicity of the trace we will work in Heisenberg
picture, shifting the action of the temporal evolution operator
on the simple expression
${a^{\scriptscriptstyle \dagger}_{f}} {a_{g}}$, thus considerably
simplifying the calculation without introducing restrictive assumptions on
the structure of ${{\varrho}^{\rm m}}$
or of the interaction.
To proceed further we introduce the following superoperators
        \[
        {\cal H}={i \over \hbar} [{H},\cdot],  \quad
        {\cal H}_0={i \over \hbar} [{H}_0 + {H}_{\rm m},\cdot],
        \quad
        {\cal V}={i \over \hbar} [{V},\cdot]       ,
        \]
acting
on the algebra generated by creation and destruction
operators.
Let us note that the
 operators
$
({a^{\scriptscriptstyle \dagger}_{h_1}})^{n_1}
({a^{\scriptscriptstyle \dagger}_{h_2}})^{n_2}
\ldots
({a^{\scriptscriptstyle \dagger}}_{h_r})^{n_r}
({a_{k_1}})^{m_1}
({a_{k_2}})^{m_2}
\ldots
(a_{k_s})^{m_s}
$
are ``eigenstates'' of the superoperator ${\cal H}_0$ with
eigenvalues
$
{i \over \hbar}
\left(
{\sum_{i=1}^{r}n_i E_{h_i} - \sum_{i=1}^{s}m_i E_{k_i}}
\right)
$, in particular:
        \[
        {\cal H}_0 {a_{{h}}} = -{i \over \hbar} {E_h} {a_{{h}}} \qquad
        {\cal H}_0 {a^{\scriptscriptstyle \dagger}_{h}} =
        +{i \over \hbar} {E_h} {a^{\scriptscriptstyle \dagger}_{h}}  .
        \]
To calculate (\ref{new}) we evaluate ${e^{{\cal H}\tau}}\left(
        {{a^{\scriptscriptstyle \dagger}_{h}}{a_{k}}}
        \right)
$
with the help of the following integral representation:
        {\openup5pt
        \begin{eqnarray}
        {e^{{\cal H}\tau}}
        \left(
        {{a^{\scriptscriptstyle \dagger}_{h}}{a_{k}}}
        \right)
        &=&
        \left(
        {e^{{\cal H}\tau}}{a^{\scriptscriptstyle \dagger}_{h}}
        \right)
               \left(
        {{e^{{\cal H}\tau}}{a_{k}}}
               \right)
        =
        \hphantom{{e^{{\cal H}\tau}}
        \left(
        {{a^{\scriptscriptstyle \dagger}_{h}}{a_{k}}}
        \right)
        }
        \nonumber \\
        & = & \!
                {\int_{-i\infty+\varepsilon}^{+i\infty +  \varepsilon}}{
        dz_1
        \over
            2\pi i
        }     \,     e^{z_1 \tau}
        \left(
        {
        {
        {
        \left(
        {{ z_1 - {\cal H}}}
        \right)
        }^{-1}}
        {a^{\scriptscriptstyle \dagger}_{h}}}
              \right)
        {\int_{-i\infty+\varepsilon}^{+i\infty +  \varepsilon}}{
        dz_2
        \over
            2\pi i
        }       \,   e^{z_2 \tau}
        \left(
        {
        {{
        \left(
        {{ z_2 - {\cal H}}}
        \right)
        }^{-1}}
        {a_{k}}}
        \right)
        .
        \label{bb7}
        \end{eqnarray}
        }
Using twice the identity:
        \begin{equation}
        {{
        \left(
        {{ z - {\cal H}}}
        \right)
        }^{-1}}
        = {{
        \left(
        {{ z - {\cal H}_0}}
        \right)
        }^{-1}}
        \left[{1+{\cal V}{{
        \left(
        {{ z - {\cal H}}}
        \right)
        }^{-1}}}\right] =
        \left[{1+{{
        \left(
        {{ z - {\cal H}}}
        \right)
        }^{-1}}{\cal V}}\right]{{
        \left(
        {{ z - {\cal H}_0}}
        \right)
        }^{-1}}
        \label{bb8}
        \end{equation}
we obtain
        \begin{equation}
        {{
        \left(
        {{ z - {\cal H}}}
        \right)
        }^{-1}}={{
        \left(
        {{ z - {\cal H}_0}}
        \right)
        }^{-1}} +{{
       \left(
        {{ z - {\cal H}_0}}
       \right)
        }^{-1}}
        {\cal T}(z){{
        \left(
        {{ z - {\cal H}_0}}
        \right)
        }^{-1}},
        \quad {\cal T}(z)\equiv
        {\cal V} + {\cal V}{{
        \left(
        {{ z - {\cal H}}}
        \right)
        }^{-1}}{\cal V} ,
        \label{211}
        \end{equation}
to be substituted in (\ref{bb7}).
Taking into account the fact that $[{H},{N}]=0$ one can see
that
the restriction to ${{{\cal H}^1_{\scriptscriptstyle F}}}$  of the
operator ${{\cal T}(z)}{a_{k}}$ has the
simple general form:
        \begin{equation}
        \left(
        {{{\cal T}(z)}{a_{k}}}
        \right)
        _{{{\cal H}^1_{\scriptscriptstyle F}}}
	=\sum_h
        T{}_{h}^{k}
        \left(
           z
        \right)
        {a_{{h}}} ,
        \label{213}
        \end{equation}
where $
T{}_{h}^{k}
\left(
   z
\right)
$ is an operator in the subspace
${{{\cal H}^0_{\scriptscriptstyle F}}}$.
This restriction is the only part of interest to us, since we are
considering a single microsystem.
One can also express $
T{}_{h}^{k}
\left(
   z
\right)
$ in terms of ${{\cal T}(z)}$ as:
        \begin{equation}
        \left[{
        \left(
        {{{\cal T}(z)} {a_{k}}}
        \right)
        {a^{\scriptscriptstyle \dagger}_{h}}
        }\right]_{{{\cal H}^0_{\scriptscriptstyle F}}} =
        T{}_{h}^{k}
        \left(
           z
        \right)
        \label{212a}
        \end{equation}
and, taking the adjoint, also
        \begin{equation}
        \left[{{a_{{h}}}
        \left(
        {{{\cal T}(z)}{a^{\scriptscriptstyle \dagger}_{k}}}
        \right)
        }\right]_{{{{\cal H}^0_{\scriptscriptstyle F}}}}     =
        {T^{\scriptscriptstyle \dagger}}{}_{h}^{k}
        \left(
        z^*
        \right)
         .   \label{212b}
        \end{equation}
Formulae
(\ref{bb8}) and
(\ref{211})    are clearly reminiscent of the usual identities
satisfied by the resolvent operator in the theory of scattering.
The mathematical framework is however quite different, since we
are now dealing with superoperators. The quantity to be related
with the usual T-matrix is the operator
$
T{}_{h}^{k}
\left(
   z
\right)
$
of (\ref{212a}), acting
in the subspace
${{{\cal H}^0_{\scriptscriptstyle F}}}$,
that is to say a second-quantized operator for the  macrosystem.
Its expectation value, which appears in the final equation
(\ref{2b12}) via the operator
${\mbox{\sf Q}}$,
may be linked to a
refractive index, often used as a phenomenological description of
the interaction of a single particle with matter (Vigu\'e, 1995),
as already mentioned in the first paragraph.
The index of refraction being an operator it would also be
possible to calculate fluctuations from the equilibrium value.
On the same footing, neglecting the incoherent contribution to the
dynamics, that is to say the last term of the Lindblad equation
(\ref{2b12}), the usual description of neutron-optics, still
based on phenomenological potentials, may be recovered (Sears,
1989).
In a future paper we intend to elucidate these possible
connections to phenomenological expressions and concrete
applications in detail.
\par
Denoting with ${\vert \lambda \rangle}\equiv
\vert 0 \rangle
\otimes{\vert \lambda \rangle}$
the basis of eigenstates of
${H}_{\rm m}$ spanning ${{{\cal H}^0_{\scriptscriptstyle F}}}$,
        $
        {H}_{\rm m}{\vert \lambda \rangle}=
        E_\lambda
        {\vert \lambda \rangle}, \label{214}
        $
we obtain the following explicit representation of $
\left(
{{{
\left({{ z - {\cal H}}}\right)
}^{-1}}{a_{k}}}
\right)
_{{{\cal H}^1_{\scriptscriptstyle F}}}$ as a mapping of
${{{\cal H}^1_{\scriptscriptstyle F}}}$
into ${{{\cal H}^0_{\scriptscriptstyle F}}}$:
        \[
        \left(
        {{{
        \left({{ z - {\cal H}}}\right)
        }^{-1}} {a_{k}}}
        \right)
        _{{{\cal H}^1_{\scriptscriptstyle F}}} =
        {
        {a_{k}}
        \over
           z+{i \over \hbar} {E_k}
        }           +
        \sum^{}_{{\lambda,\lambda' \atop  f}}
        {
        {\vert \lambda' \rangle}
        {\langle
        \lambda'
        \vert
        T{}_{f}^{k}
        \left(
           z
        \right)
        \vert
                \lambda
        \rangle}
        {\langle \lambda \vert}
        {a_{{f}}}
        \over
        \left(
        {z + {i \over \hbar} {E_k}}
        \right)
        \left(
        {z - {i \over \hbar}
        \left({{E_{{\lambda}'}} - {E_{{\lambda}}} - {E_f}}\right)
        }
        \right)
        }  .
        \]
Since $
\left(
{{{
\left({{ z^* - {\cal H}}}\right)}^{-1}} {a_{k}}}
\right)
^{\scriptscriptstyle \dagger}= {{
\left(
{{ z - {\cal H}}}
\right)
}^{-1}}{a^{\scriptscriptstyle \dagger}_{k}}$
and by (\ref{25})
one has easily:
        {\openup2pt
        \begin{eqnarray}
        \label{217}
        &&
        \!\!\!\!\!\!\!\!\!  
        {\hbox{\rm Tr}}_{
        {{{\cal H}_{\scriptscriptstyle F}}}
        }
        \left[{
        \left(
        {
        {{
        \left({{ z_1 - {\cal H}}}\right)}^{-1}}
        {a^{\scriptscriptstyle \dagger}_{h}}}
              \right)
        \left(
        {{{
        \left({{ z_2 - {\cal H}}}\right)
        }^{-1}}        {a_{k}}}
        \right)
        {\varrho}(t)}\right]=
        \nonumber \\
        &&
        \!\!\!\!\!\!\!\!\!  
        \hphantom{=}
        =
        {
        {\varrho}_{kh}^{(1)}(t)
        \over
        \left(
        {z_1 - {i \over \hbar} {E_h}}
        \right)
        \left(
        {z_2 + {i \over \hbar} {E_k}}
        \right)
        }
        \nonumber \\
        &&
        \!\!\!\!\!\!\!\!\!  
        \hphantom{==}
        \mbox{}+
        \sum_{{\lambda,\lambda'\atop g  }}
        {
        1
        \over
        {z_2 + {i \over \hbar} {E_k}}
        }
        {\varrho}_{kg}^{(1)}(t)
                {
        {\langle
        \lambda
        \vert
        {T^{\scriptscriptstyle \dagger}}{}_{g}^{h}
        \left(
        z_1^*
        \right)
        \vert
                \lambda'
        \rangle}
        \langle\lambda'\vert{\varrho^{\rm m}(t)}\vert\lambda\rangle
        \over
        \left(
        {z_1 - {i \over \hbar} {E_h}}
        \right)
                \left(
                {z_1 + {i \over \hbar}
                \left({{E_{{\lambda}'}} - {E_{{\lambda}}} - {E_g}}
                \right)}
                \right)
        }
        \nonumber \\
        &&
        \!\!\!\!\!\!\!\!\!  
        \hphantom{==}
        \mbox{}+
        \sum_{{\lambda,\lambda'\atop f  }}
        {
        {\langle
        \lambda'
        \vert
        T{}_{f}^{k}
        \left(
           z_2
        \right)
        \vert
                \lambda
        \rangle}
        \langle
        \lambda\vert{\varrho^{\rm m}(t)}\vert\lambda'\rangle
        \over
        \left(
        {z_2 + {i \over \hbar} {E_k}}
        \right)
        \left(
        {z_2 - {i \over \hbar}
        \left({{E_{{\lambda}'}} - {E_{{\lambda}}} - {E_f}}\right)
        }
        \right)
        }
        {\varrho}_{fh}^{(1)}(t)
        {
        1
        \over
         z_1 - {i \over \hbar} {E_h}
        }
        \nonumber \\
        &&
        \!\!\!\!\!\!\!\!\!  
        \hphantom{==}
        \mbox{}+
        \sum_{{\lambda,\lambda' ,\lambda''\atop f,g }}
        {
        \langle\lambda''\vert
        T{}_{f}^{k}
        \left(
           z_2
        \right)
        \vert\lambda\rangle
        \over
        \left(
        {z_2 + {i \over \hbar} {E_k}}
        \right)
        \left(
        {z_2 - {i \over \hbar}
        \left({{E_{{\lambda}''}} - {E_{{\lambda}}} -
        {E_f}}\right)
        }
        \right)
        }
        \nonumber \\
        &&
        \!\!\!\!\!\!\!\!\!  
        \hphantom{==+..\sum_{{\lambda,\lambda' ,\lambda''\atop f,g }}}
        \mbox{}
        \times
        \langle
        \lambda\vert{\varrho^{\rm m}(t)}
        \vert\lambda'\rangle
        {
        {\langle
        \lambda'
        \vert
        {T^{\scriptscriptstyle \dagger}}{}_{g}^{h}
        \left(
        z_1^*
        \right)
        \vert
                \lambda''
        \rangle}
        \over
        \left(
        {z_1 - {i \over \hbar} {E_h}}
        \right)
        \left(
        {z_1 + {i \over \hbar}
        \left({{E_{{\lambda}''}} - {E_{{\lambda}'}} -
        {E_g}}\right)
        }
        \right)
        }
        {\varrho}_{fg}^{(1)}(t).
        \end{eqnarray}
        }
Since these expressions will be considered for values of the
complex variables $z,z_1,z_2$ of the form
$iy+\varepsilon$ we can replace
in  (\ref{217})
${E_h}\rightarrow {E_h}-i\hbar\eta$,
${E_k} \rightarrow {E_k} +i\hbar\eta$,
${E_f} \rightarrow {E_f} +2i\hbar\eta$,
${E_g} \rightarrow {E_g} -2i\hbar\eta$, $\varepsilon>\eta> 0$,
without
introducing singularities and obtaining expressions that depend
smoothly on the parameter $\eta$ and yield (\ref{217}) in the limit
$\eta \rightarrow 0$.
Let us consider the expression:
        \[
        {{\mbox{\bf Q}}}^{\scriptscriptstyle \dagger}_{gh}(\tau,\eta)
        =
        {\int_{-i\infty+\varepsilon}^{+i\infty +  \varepsilon}}{
        dz
        \over
            2\pi i
        }     \,     e^{
        \left(
        {z-{i \over \hbar} {E_k}+\eta}
        \right)
         \tau}
        \sum_{{\lambda,\lambda'  }}
                {
                {\langle
        \lambda
        \vert
        {T^{\scriptscriptstyle \dagger}}{}_{g}^{h}
        \left(
        z^*
        \right)
        \vert
                \lambda'
        \rangle}
        \langle\lambda'\vert{\varrho^{\rm m}(t)}\vert\lambda\rangle
        \over
        \left(
        {z - {i \over \hbar} {E_h}-\eta}
        \right)
        \! \!
        \left(
        {z + {i \over \hbar}
        \left({{E_{{\lambda}'}} - {E_{{\lambda}}} -  {E_g}}\right)
        -2\eta}
        \right)
        };
        \]
in the integration over $z$ we will distinguish two different
kinds of contributions; the first one due to the denominators and
strongly dependent on the indexes $g,h$, the second one due to the
singularities of $
        {T^{\scriptscriptstyle \dagger}}{}_{g}^{h}
        \left(
        z^*
        \right)
$ that are poles on
the imaginary axis:
        \[
        {{\mbox{\bf Q}}}^{{\scriptscriptstyle \dagger}}_{gh}(\tau,\eta )=
        {{\mbox{\bf Q}}}^{{\scriptscriptstyle \dagger}}_{1gh}(\tau,\eta ) +
        {{\mbox{\bf Q}}}^{{\scriptscriptstyle \dagger}}_{2gh}(\tau,\eta )
        \]
We obtain:
        {\openup5pt
        \begin{eqnarray}
        \label{219}
        &&
        \!\!\!\!\!\!
        {{\mbox{\bf Q}}}^{{\scriptscriptstyle \dagger}}_{1gh}(\tau,\eta) =
        \sum_{{\lambda,\lambda'  }}
        {
        e^{{i \over \hbar}
        \left(
        {{E_h}-{E_k}}
        \right)
        \tau+2\eta \tau}
        \over
        {i \over \hbar}
        \left(
        {{E_{{\lambda}'}}+{E_h}-{E_{{\lambda}}}-{E_g}}
        \right)
         -\eta}
        \langle
        {\lambda}
        \vert
        {T^{\scriptscriptstyle \dagger}}{}_{g}^{h}
        \left(
        {-{i \over \hbar} {E_h}+\eta}
        \right)
        \vert
        {\lambda'}
        \rangle
        {{\varrho}^{\rm m}_{\lambda'\lambda}(t)}
        \nonumber \\
        &&
        \!\!\!\!\!\!
        \mbox{}+
        \sum_{{\lambda,\lambda'  }}
        {
        e^{-{i \over \hbar}
        \left(
        {{E_{{\lambda}'}}+{E_k}-{E_{{\lambda}}}-{E_g}}
        \right)
        \tau   +3\eta \tau  }
        \over
        {-{i \over \hbar}
        \left(
        {{E_{{\lambda}'}}+{E_h}-{E_{{\lambda}}}-{E_g}}
        \right)
           +\eta    }
        }
        \langle
        {\lambda}
               \vert
        {T^{\scriptscriptstyle \dagger}}{}_{g}^{h}
        \left(
        {i \over \hbar}
        \left(
        {{E_{{\lambda}'}} -{E_{{\lambda}}}-{E_g}}
        \right)
           +2\eta
        \right)
                    \vert
           {\lambda'}
                         \rangle
        {{\varrho}^{\rm m}_{\lambda'\lambda}(t)}
        \nonumber \\
        &&
        \!\!\!\!\!\!
        =
        \sum_{{\lambda,\lambda'  }}
        e^{{i \over \hbar}
        \left(
        {{E_h}-{E_k}}
        \right)
        \tau+2\eta \tau}
        {
        1 - e^{-{i \over \hbar}
        \left(
        {{E_{{\lambda}'}}+{E_h}-{E_{{\lambda}}}-{E_g}}
        \right)
        \tau   +\eta \tau  }
        \over
        {i \over \hbar}
        \left(
        {{E_{{\lambda}'}}+{E_h}-{E_{{\lambda}}}-{E_g}}
        \right)
         -\eta }
        \langle
        {\lambda}
               \vert
        {T^{\scriptscriptstyle \dagger}}{}_{g}^{h}
        \left(
        {-{i \over \hbar} {E_h}+\eta}
        \right)
                    \vert
        {\lambda'}
                         \rangle
        {{\varrho}^{\rm m}_{\lambda'\lambda}(t)}
        \nonumber
        \\
        &&
        \mbox{}+
        \sum_{{\lambda,\lambda'  }}
        e^{-{i \over \hbar}
        \left(
        {{E_{{\lambda}'}}+{E_k}-{E_{{\lambda}}}-{E_g}}
        \right)
        \tau   +3\eta \tau  }
        \nonumber \\
        &&
        \hphantom{\mbox{}+\sum}
        \times               
        {
             \langle
         {\lambda}
                  \vert
        {T^{\scriptscriptstyle \dagger}}{}_{g}^{h}
        \left(
        {+{i \over \hbar}
        \left(
        {{E_{{\lambda}'}}
        -{E_{{\lambda}}}-{E_g}}
        \right)
           +2\eta}
        \right)
           -
        {T^{\scriptscriptstyle \dagger}}{}_{g}^{h}
        \left(
        {-{i \over \hbar} {E_h}+\eta}
        \right)
                       \vert
           {\lambda'}
                            \rangle
	\over
        \left(
	{-{i \over \hbar}
        \left(
        {{E_{{\lambda}'}}-{E_{{\lambda}}}-{E_g}}
        \right)
           +2\eta   }
	\right)
	-
	\left( + {i \over \hbar}  E_h + \eta \right)
        }  
        {{\varrho}^{{\rm m}}_{\lambda'\lambda}(t)}
        .
        \end{eqnarray}
        }
If we choose a time scale, dependent on the properties of the
statistical operator, such that
        \begin{equation}
        \left|
        {{E_{{\lambda}'}}+{E_h}-{E_{{\lambda}}}-{E_g}}
        \right|
        {\tau \over \hbar} \ll
        1 , \label{219a}
        \end{equation}
we can simply retain in the first factor the contribution linear
in $\tau$, which amounts to
        \[
        \tau        \sum_{{\lambda,\lambda'  }}
        \langle
        {\lambda}
               \vert
        {T^{\scriptscriptstyle \dagger}}{}_{g}^{h}
        \left(
        {-{i \over \hbar} {E_h}+\eta}
        \right)
                    \vert
        {\lambda'}
                           \rangle
        \langle\lambda'\vert{\varrho^{\rm m}(t)}\vert\lambda\rangle .
        \]
The second term is a superposition of a huge set of exponentials
$e^{-{i \over \hbar}
\left(
{{E_{{\lambda}'}}+{E_k}-{E_{{\lambda}}}-{E_g}}
\right)
\tau}$ with amplitudes
        \[
        {
           \langle
        {\lambda}
                  \vert
        {T^{\scriptscriptstyle \dagger}}{}_{g}^{h}
        \left(
        {+{i \over \hbar}
        \left(
        {{E_{{\lambda}'}} -{E_{{\lambda}}}-{E_g}}
        \right)
           +2\eta}
        \right)
           -
        {T^{\scriptscriptstyle \dagger}}{}_{g}^{h}
        \left(
        {-{i \over \hbar} {E_h}+\eta}
        \right)
                       \vert
           {\lambda'}
                            \rangle
        \over
        {-{i \over \hbar}
        \left(
        {{E_{{\lambda}'}}+{E_h}-{E_{{\lambda}}}-{E_g}}
        \right)
         +\eta}
        }
        \]
that are slowing varying on a range $\sigma$ of the variable
${{1 \over \hbar}
\left(
{{E_{{\lambda}'}}+{E_k}-{E_{{\lambda}}}-{E_g}}
\right)
}$, as long as
$\eta$ is large with respect to the spacing between the values of
this variable; then the second term of (\ref{219}) is negligible for
$\tau\gg  {1\over\sigma}$, where ${1\over\sigma}$ may be identified
with the
correlation time for the macrosystem; we are thus working on a
time scale long enough to ignore fluctuations from the
non-perturbed state for the macrosystem.
Since by  (\ref{211})
${{\cal T}(z)}$
has
poles on the imaginary axis at the points ${i \over \hbar}
\left(
{\xi_\lambda -\xi_{\lambda'}}
\right)
$, $\xi_\lambda$ being the eigenvalues of ${H}$,
and therefore by  (\ref{212b}) also $
        {T^{\scriptscriptstyle \dagger}}{}_{l}^{k}
        \left(
        z^*
        \right)
$
has such poles,
as we did
before we shall assume that the superposition of this huge set of
contributions makes
${{\mbox{\bf Q}}}^{{\scriptscriptstyle \dagger}}_{2gh}(\tau,\eta )$
negligible if $\tau\gg {1\over\sigma}$; then we have
the simple asymptotic result:
        \begin{equation}
        \label{222}
        {{\mbox{\bf Q}}}^{{\scriptscriptstyle \dagger}}_{gh}(\tau,\eta )
         =
        \tau
        {\hbox{\rm Tr}}_{{{{\cal H}_{\scriptscriptstyle F}}}}
        \left[{{a_{{g}}}
        \left(
        {\cal T}
        \left({{i \over \hbar} {E_h}+\eta}\right)
         {a^{\scriptscriptstyle \dagger}_{h}}
              \right)
        {{\varrho}^{\rm m}(t)}
        }\right]
        \qquad
        {1\over\sigma} \ll \tau \ll \tau_1  \quad \eta \gg \delta,
        \end{equation}
where $\delta$ is the spacing between the poles of
$
T(z)
$ and $\tau_1$ represents the typical variation time
inside the reduced description; $\tau_1$ must be large enough, i.e.
the reduced dynamics must be slow enough to justify (\ref{219a}).
Correspondingly the statistical operator of the microsystem
must be such that:
        \begin{equation}
        {\varrho}_{gf}^{(1)} \simeq 0  \quad
        {{\rm \hbox{\quad if \quad}}}
        \quad
        {
        {E_g}- {E_f}
        \over
                \hbar
        }
        \simeq {1\over \tau_1} \label{223}
        \end{equation}
and the statistical operator ${{\varrho}^{\rm m}(t)}$ must
be close enough to
an equilibrium statistical operator:
        \begin{equation}
        {{\varrho}^{\rm m}(t)}_{\lambda\lambda'}\simeq 0
        \quad
        {{\rm \hbox{\quad if \quad}}}
        \quad
        {
        E_\lambda
        - {E_{\lambda'}}
        \over
                \hbar
        }
        \geq {1\over\tau_1}  . \label{224}
        \end{equation}
Let us now concentrate on the expression
        {\openup5pt
        \begin{eqnarray*}
        {\mbox{\bf L}}_{kfgh}(\tau,\eta ) 
	\!\! 
        &=&
	\!\!
        {\int_{-i\infty+\varepsilon}^{+i\infty +  \varepsilon}}{
        dz_1
        \over
            2\pi i
        }     \,     
        {\int_{-i\infty+\varepsilon}^{+i\infty +  \varepsilon}}{
        dz_2
        \over
            2\pi i
        } \, e^{
        \left(
        {z_1+z_2}
        \right)
         \tau}
        \,
        \sum_{{\lambda,\lambda'\atop\lambda'' }}
        {
        \langle\lambda''\vert
        T{}_{f}^{k}
        \left(
           z_2
        \right)
        \vert\lambda\rangle
        \over
        \left(
        {z_2 + {i \over \hbar} {E_k}}
        \right)
        \left(
        {z_2 - {i \over \hbar}
        \left({{E_{{\lambda}''}} - {E_{{\lambda}}} -
        {E_f}}\right)
        }
        \right)
        }
        \\
        &&
        \times
        \langle
        \lambda\vert{\varrho^{\rm m}(t)}\vert\lambda'\rangle
        {
        {\langle
        \lambda'
        \vert
        {T^{\scriptscriptstyle \dagger}}{}_{g}^{h}
        \left(
        z_1^*
        \right)
        \vert
                \lambda''
        \rangle}
        \over
        \left(
        {z_1 - {i \over \hbar} {E_h}}
        \right)
        \left(
        {z_1 + {i \over \hbar}
        \left({{E_{{\lambda}''}} - {E_{{\lambda}'}} -
        {E_g}}\right)
        }
        \right)
        }
        ,
        \end{eqnarray*}
        }
by a similar procedure, neglecting the singularities of
$
T(z)
$
and taking into account the slow variability of
$
        T{}_{f}^{k}
        \left(
           iy+\eta
        \right)
$
one has:
        \begin{eqnarray*}
        &&
         {\mbox{\bf L}}_{kfgh}(\tau,\eta )=
        \sum_{{\lambda,\lambda',\lambda''  }}
        {
        \hbar^2
        \over
        \left(
        {{E_h}+{E_{{\lambda}''}}-{E_g}-{E_{{\lambda}'}} +i\hbar\eta}
        \right)
        \left(
        {{E_k}+{E_{{\lambda}''}}-{E_f}-{E_{{\lambda}}}    -i\hbar\eta}
        \right)
        }
        \\
        &&
        \Biggl\{
        e^{{i \over \hbar}
        \left(
        {{E_h}-{E_k}}
        \right)
        \tau+2\eta \tau}
        \langle
        {\lambda''}
        \vert
        {
                T{}_{f}^{k}
        \left(
        {{-{i \over \hbar}
        {E_k}+\eta}}
        \right)
        }
        \vert
        {\lambda}
        \rangle
        {{\varrho}^{\rm m}_{\lambda\lambda'}(t)}
        \langle
        {\lambda'}
        \vert
        {T^{\scriptscriptstyle \dagger}}{}_{g}^{h}
        \left(
        {-{i \over \hbar} {E_h}+\eta}
        \right)
        \vert
        {\lambda''}
        \rangle
        \\
        &&
        {\>}+
        e^{-{i \over \hbar}
        \left(
        {{E_f}-{E_g}}
        \right)
        \tau+4\eta \tau}
        \langle
        {\lambda''}
        \vert
        {
                T{}_{f}^{k}
        \left(
        {{
        {i \over \hbar}
                \left(
                {{E_{\lambda''}}-E_\lambda-{E_f}}
                \right)
                 +2\eta}}
        \right)
        }
        \vert
                {\lambda}
        \rangle
                {{\varrho}^{\rm m}_{\lambda\lambda'}(t)}
        \\
        &&
        \hphantom{\mbox{}+{}}
        \times
        \langle
        {\lambda'}
        \vert
        {T^{\scriptscriptstyle \dagger}}{}_{g}^{h}
        \left(
        {{{i \over \hbar}
        \left(
        {{E_{\lambda''}}     -{{E_{\lambda'}}}-{E_g}}
        \right)
        +2\eta}}
        \right)
        \vert
        {\lambda''}
        \rangle
        \\
        &&
        {\>}-
        e^{{i \over \hbar}
        \left(
        {{E_h}+
        {E_{\lambda''}}
        -{E_f}-
        E_\lambda
        }
              \right)
        \tau+3\eta \tau}
        \langle
        {\lambda''}
        \vert
        {
        T{}_{f}^{k}
        \left(
        {{{i \over \hbar}
                \left(
                {{E_{\lambda''}}-E_\lambda-{E_f}}
                \right)
                 +2\eta}
                }
        \right)
        }
        \vert
                {\lambda}
        \rangle
                {{\varrho}^{\rm m}_{\lambda\lambda'}(t)}
        \\
        &&
        \hphantom{\mbox{}+{}}
        \times
        \langle
        \lambda'
        \vert
        {T^{\scriptscriptstyle \dagger}}{}_{g}^{h}
        \left(
        {-{i \over \hbar} {E_h}+\eta}
        \right)
        \vert
        {\lambda''}
        \rangle
        \\
        &&
        {\>}-
        e^{
	{i \over \hbar}
        \left(
        {E_g}+
        {E_{\lambda'}}
        -{E_k}-
        {E_{\lambda''}}
              \right)
        \tau+3\eta \tau}
        \langle
        {\lambda''}
        \vert
        {
        T{}_{f}^{k}
        \left(
        {{-{i \over \hbar}{E_k}+\eta}}
        \right)
        }
        \vert
        {\lambda}
        \rangle
        {{\varrho}^{\rm m}_{\lambda\lambda'}(t)}
        \\                                      
         &&
        \hphantom{\mbox{}+{}}
        \times
        \langle
        {\lambda'}
        \vert
        {T^{\scriptscriptstyle \dagger}}{}_{g}^{h}
        \left(
        {{{i \over \hbar}
        \left(
        {{{E_{\lambda''}}}-{{E_{\lambda'}}}-
        {E_g}}
              \right)
        +2\eta}}
        \right)
        \vert
        {\lambda''}
        \rangle
        \Biggr\}
        .
        \end{eqnarray*}
Arguing as before we can extract from this expression the
dominant part:
        \begin{eqnarray}
        \label{relevant}
        &&
        \sum^{}_{{\lambda,\lambda',\lambda''  }}
        \hbar^2
        {
        \langle
        {\lambda''}
        \vert
        {
                T{}_{f}^{k}
        \left(
        {{-{i \over \hbar}{E_k}+\eta}}
        \right)
        }
        \vert
        {\lambda}
        \rangle
        {{\varrho}^{\rm m}_{\lambda\lambda'}(t)}
        \langle
        {\lambda'}
        \vert
        {T^{\scriptscriptstyle \dagger}}{}_{g}^{h}
        \left(
        {-{i \over \hbar} {E_h}+\eta}
        \right)
        \vert
        {\lambda''}
        \rangle
        \over
        \left(
        {{E_h}+{E_{{\lambda}''}}-{E_g}-{E_{{\lambda}'}} +i\hbar\eta}
        \right)
        \left(
        {{E_k}+{E_{{\lambda}''}}-{E_f}-{E_{{\lambda}}} -i\hbar\eta}
        \right)
        }
        \nonumber \\
        &&
        \hphantom{\sum^{}_{{\lambda,\lambda',\lambda''  }}}\times\biggl[
        e^{{i \over \hbar}
        \left(
        {{E_h}-{E_k}}
        \right)
        \tau+2\eta \tau}
        -
        e^{{i \over \hbar}
        \left(
        {{E_h}+{{E_{\lambda''}}}-{E_f}-{E_\lambda}}
        \right)
        \tau+3\eta \tau}
        \nonumber \\
        &&
        \hphantom{\sum^{}_{{\lambda,\lambda',\lambda''  }}\times\biggl[}
        -e^{{i \over \hbar}
        \left(
        {{E_g}+{{E_{\lambda'}}}-{E_k}-{{E_{\lambda''}}}}
        \right)
        \tau+3\eta \tau}
        +
        e^{{i \over \hbar}
        \left(
        {{E_g}-{E_f}}
        \right)
        \tau+4\eta \tau}
        \biggr]
        .
        \end{eqnarray}
The evaluations  (\ref{222}) and (\ref{relevant}) hold for a finite
value of
the parameter $\eta$; in the limit $\eta \to 0$ singularities
arise in these expressions that would be compensated by
singularities coming from the neglected contributions: the
splitting of
${{\mbox{\bf Q}}}^{\scriptscriptstyle \dagger}_{gh}(\tau,\eta )$ and
${\mbox{\bf L}}_{kfgh}(\tau,\eta )$ into a relevant and a negligible part
becomes therefore meaningless. For a finite confined system this
treatment unavoidably relies on an approximation. The situation
can be improved considering the limit of no confinement: then the
set of eigenvalues
$
\left \{{{E_g}}\right \}$ and
$\left \{{{E_\lambda}}\right \}$ becomes a continuum; expressions
of the
form $
\langle
{\lambda}
  \vert
        T{}_{f}^{g}
        \left(
        z
        \right)
  \vert
{\lambda'}
  \rangle
$ become analytic
functions for $Re\, z >0$, having a cut on the imaginary axis and
the existence of the limit $\delta \to 0$ can be reasonably
assumed. The analytic continuation across the cut can be
considered and one can assume that the singularities of this
continuation are located in the left half-plane far enough from
the imaginary axis to give contributions that rapidly decay for
$\tau\gg{1\over\sigma}$, thus providing the precise reason that makes
the previously considered terms indeed negligible. In this way a
further simplification of (\ref{relevant}) becomes clear: if the
sum over
${{E_{\lambda''}}}$ (or ${{E_{\lambda'}}}$) is eventually replaced by an
integral and the integration path shifted inside the complex
${{E_{\lambda''}}}$ plane, the contribution of the term
$e^{{i \over \hbar}
\left(
{{E_h}+{{E_{\lambda''}}}-{E_f}-{E_\lambda}}
\right)
\tau+3\eta \tau}$
can be calculated shifting the integration path for  ${{E_{\lambda''}}}$
in the upper half-plane; then the only contribute of the
singularity
${
1
\over
 {E_k}+{E_{{\lambda}''}}-{E_f}-{E_{{\lambda}}}    -i\hbar\eta
}$ lying in the upper half-plane must be considered, so that
replacing
${{E_{\lambda''}}}$
by
${{E_{\lambda''}}}=({E_\lambda}+{E_f}-{E_k}+i\hbar\eta)$ the term becomes
$e^{{i \over \hbar}
\left(
{{E_h}-{E_k}}
\right)
\tau+2\eta \tau}$.
Similarly
$e^{{i \over \hbar}
\left(
{{E_g}+{{E_{\lambda'}}}-{E_k}-{{E_{\lambda''}}}}
\right)
\tau+3\eta \tau}$
replacing
${{E_{\lambda''}}}=({{E_{\lambda'}}}+{E_g}-{E_h}-i\hbar\eta)$ becomes
$e^{{i \over \hbar}
\left(
{{E_h}-{E_k}}
\right)
\tau+2\eta \tau}$.
We thus obtain for the
square bracket in  (\ref{relevant}):
        \[
        \left[{
        e^{{i \over \hbar}
        \left(
        {{E_g}-{E_f}}
        \right)
        \tau+4\eta \tau}
        -
        e^{{i \over \hbar}
        \left(
        {{E_h}-{E_k}}
        \right)
        \tau+2\eta \tau}
        }\right]
        \simeq
        2\eta \tau +{i \over \hbar}
        \left(
        {{E_g}-{E_f}+{E_k}-{E_h}}
        \right)
        \tau
        \]
Keeping $\eta$ finite and appealing to (\ref{223})
we are led to keep only the first contribution. As
mentioned previously the limit $\eta \to 0$ cannot be taken at
any arbitrary step of the calculation, which
in its intermediate steps
essentially relies
upon the finiteness of $\eta$ [see (\ref{222})];
 anyway it is to be expected that this limit can be considered
 after taking the continuous limit on the set
 $\{ E_{\alpha}\}$.
By this systematic asymptotic evaluation of
(\ref{217})
we come to the following:
        {\openup5pt
        \begin{eqnarray*}
        {\varrho}_{kh}^{(1)}(t+\tau)
        &=&
        {\hbox{\rm Tr}}_{{{\cal H}_{\scriptscriptstyle
        F}}}
        \left[{{e^{{\cal H}\tau}}
        \left(
        {{a^{\scriptscriptstyle \dagger}_{h}}{a_{k}}}
        \right)
        {{\varrho}}(t)}\right]
        \\
        &=&
        {\varrho}_{kh}^{(1)}(t)
        -{i \over \hbar}\tau
        \left(
        {{E_k}-{E_h}}
        \right)
          {\varrho}_{kh}^{(1)}(t)
        \\
        &&
        \mbox{}+
        \tau\sum_g         {\varrho}_{kg}^{(1)}(t)
        {\hbox{\rm Tr}}_{{{{\cal H}_{\scriptscriptstyle
        F}}}}
        \left[{{a_{{g}}}
        \left(
        {{\cal T}
        \left({{i \over \hbar} {E_h}+\eta}\right)
        {a^{\scriptscriptstyle \dagger}_{h}}
        }
              \right)
        {{\varrho}^{\rm m}(t)}
        }\right]
        \\
        &&
        \mbox{}+        \tau\sum_g
        {\hbox{\rm Tr}}_{{{{\cal H}_{\scriptscriptstyle
        F}}}}
        \left[{
        \left(
        {\cal T}
        \left({-{i \over \hbar} {E_k} +\eta}\right)
           {a_{k}}
        \right)
        {a^{\scriptscriptstyle \dagger}_{f}}
        {{\varrho}^{\rm m}(t)}
        }\right]
        {\varrho}_{fh}^{(1)}(t)
        \\
        &&
        \mbox{}+ 2\eta \hbar^2 \tau
        \sum^{}_{{\lambda,\lambda',\lambda''  \atop f,g}}
        {\varrho}_{fg}^{(1)}(t)
        {
        \langle
        {\lambda''}
        \vert
                T{}_{f}^{k}
        \left(
        {{-{i \over \hbar}{E_k}+\eta}}
        \right)
        \vert
        {\lambda}
        \rangle
        \over
        \left(
        {{E_k}+{E_{{\lambda}''}}-{E_f}-{E_{{\lambda}}}    -i\hbar\eta}
        \right)
        }
        \langle\lambda\vert{\varrho^{\rm m}(t)}\vert\lambda'\rangle
        \\
        &&
        \hphantom{\mbox{}+{}}\times
        {
        \langle
        {\lambda'}
        \vert
        {T^{\scriptscriptstyle \dagger}}{}_{g}^{h}
        \left(
        {-{i \over \hbar} {E_h}+\eta}
        \right)
        \vert
        {\lambda''}
        \rangle
        \over
        \left(
        {{E_h}+{E_{{\lambda}''}}-{E_g}-{E_{{\lambda}'}} +i\hbar\eta}
        \right)
        }
        ,
        \end{eqnarray*}
        }
and recalling (\ref{new})
        \begin{eqnarray}
        \label{2b10}
        &&
        \!   \! \! \!        \!        \!
        \!   \! \! \!        \!        \!
        \!   \! \! \!        \!        \!
        {
        d {\varrho}_{kh}^{(1)}(t)
        \over
        dt
        }
        =
        \\
        &&
        \!   \! \! \!        \!        \!
        \!   \! \! \!        \!        \!
        \!   \! \! \!        \!        \!
        =
        -{i \over \hbar}
        \left(
        {{E_k}-{E_h}}
        \right)
          {\varrho}_{kh}^{(1)}(t)
        +
        {1 \over \hbar}
        \sum_g  {\varrho}_{kg}^{(1)}(t) {\mbox{\sf Q}}^{\scriptscriptstyle
         \dagger}_{gh}
        +
        {1 \over \hbar}
        \sum_f  {\mbox{\sf Q}}_{kf} {\varrho}_{fh}^{(1)}(t)
        +
        {1 \over \hbar}
        \sum_{fg}   {\varrho}_{fg}^{(1)}(t)
       {\mbox{\bf L}}_{kfgh},
        \nonumber
        \end{eqnarray}
which
shows the structure of the generator ${\cal L}$,
where
        {\openup3pt
        \begin{eqnarray*}
        \!\!\!\!\!\!\! \!\!\!
        {\mbox{\sf Q}}_{kf}
        &=&
        \hbar{\hbox{\rm Tr}}_{{{{\cal H}_{\scriptscriptstyle F}}}}
        \left[{
        \left(
        {\cal T}
        \left({-{i \over \hbar} {E_k} +\eta}\right)
           {a_{k}}
        \right)
        {a^{\scriptscriptstyle \dagger}_{f}}
        {{\varrho}^{\rm m}(t)}
        }\right]
        \\
        \!\!\!\!\!\!\!\!\!\!  
        {\mbox{\sf Q}}^{\scriptscriptstyle \dagger}_{gh}
        &=&
        \hbar
        {\hbox{\rm Tr}}_{{{{\cal H}_{\scriptscriptstyle
        F}}}}
        \left[{{a_{{g}}}
        \left(
        {{\cal T}
        \left({{i \over \hbar} {E_h}+\eta}\right)
        {a^{\scriptscriptstyle \dagger}_{h}}
        }
              \right)
        {{\varrho}^{\rm m}(t)}
        }\right]
        \\
        \!\!\!\!\!\!\!\!\!\!  
       {\mbox{\bf L}}_{kfgh}
        &=&
        2\eta \hbar^3   \!
        \sum^{}_{{\lambda,\lambda'\atop \lambda''}}
        {
        \langle
        {\lambda''}
        \vert
        T{}_{f}^{k}
        \left(
        {{-{i \over \hbar}{E_k}+\eta}}
        \right)
        \vert
        {\lambda}
        \rangle
        {\varrho}^{\rm m}_{\lambda\lambda'}(t)
        \langle
        {\lambda'}
        \vert
        {T^{\scriptscriptstyle \dagger}}{}_{g}^{h}
        \left(
        {-{i \over \hbar} {E_h}+\eta}
        \right)
        \vert
        {\lambda''}
        \rangle
        \over
        \left(
        {{E_k}+{E_{{\lambda}''}}-{E_f}-{E_{{\lambda}}}    -i\hbar\eta}
        \right)
	\!
        \left(
        {{E_h}+{E_{{\lambda}''}}-{E_g}-{E_{{\lambda}'}} +i\hbar\eta}
        \right)
        }.
        \end{eqnarray*}
        }
By the splitting:
        \[
        {\mbox{\bf L}}_{kfgh} =
        \sum^{}_{{\xi,\lambda  }}
        \pi_{\xi}
        \left(
        {{\mbox{\sf L}}_{\lambda\xi}}
        \right)
        _{kf}
        \left(
        {{{\mbox{\sf L}}_{\lambda\xi}}}
        \right)
        ^*_{hg} \ ,
        \]
where
          {\openup5pt
          \begin{eqnarray}
          \!\!\!\!\!\!\!\!\!
          \left(
          {{\mbox{\sf L}}_{\lambda\xi}}
          \right)
          _{kf}
          &=&
          \label{2b11}
          \\
          &=&
          \sqrt{2\eta\hbar^3}
          {\langle
          \lambda
          \vert
          \left[{
          \left(
          {{\cal T}
          \left({-{i \over \hbar} {E_k} +\eta}\right)
             {a_{k}}}
          \right)
          {a^{\scriptscriptstyle \dagger}_{f}}
          }\right]
          \left(
          {{E_k}+{E_{{\lambda}}}-{E_f}-{H}_{\rm m} -i\hbar\eta}
          \right)
          ^{-1}
          \vert
          {\xi(t)}
          \rangle}
          \nonumber
          \end{eqnarray}
          }
$\xi(t)$ being a complete system of
eigenvectors of
${{\varrho}^{\rm m}(t)}$,
$({{\varrho}^{\rm m}(t)}=\sum_{\xi(t)}
\pi_{\xi(t)}
\vert {{\xi(t)}} \rangle
{\langle \xi(t) \vert})$,
and introducing in ${{\cal H}^{(1)}}$   the operators  $\mbox{\sf Q},
{\mbox{\sf L}}_{\lambda\xi}$:
        \[
        \langle
        {k}
        \vert
        {\mbox{\sf Q}}
        \vert
        {f}
        \rangle
        =\mbox{\sf Q}_{kf}
        \quad
        ,
        \quad
        \langle
        {k}
        \vert
        {{\mbox{\sf L}}_{\lambda\xi}}
        \vert
        {f}
        \rangle
        =
         {\bigl( {{\mbox{\sf L}}_{\lambda\xi}} \bigr)}_{kf}
        \]
we get the desired expression:
        \begin{equation}
        {
        d {\varrho}^{(1)}(t)
        \over
                      dt
        }
        =
        -{i \over \hbar}
        \left[{\mbox{\sf H},{\varrho}^{(1)}(t)}\right]
        +  {1\over 2\hbar}
        \left \{
        {
        \left(
        {\mbox{\sf Q}}+ {\mbox{{\sf Q}}^{\scriptscriptstyle \dagger}}
        \right)
        ,{\varrho}^{(1)}(t)}
                \right \}
        +
        {1 \over \hbar}
        \sum^{}_{{\xi,\lambda  }}
        \pi_{\xi}
        {\mbox{\sf L}}_{\lambda\xi} {\varrho}^{(1)}(t)
        {{\mbox{\sf L}}_{\lambda\xi}^{\scriptscriptstyle \dagger}}\ ,
        \label{2b12}
        \end{equation}
where
        \[
        \mbox{\sf H}=\mbox{\sf H}_0 + {i\over 2}
        \left(
        {\mbox{\sf Q}- {\mbox{\sf Q}}^{\scriptscriptstyle \dagger}}
        \right)
         .
        \]
There is still
one most important  check to be done,
that is to say we have to verify that
conservation of the trace of the
statistical operator has not been affected by the way we have
extracted the completely positive evolution  (\ref{2b12}) from the
Hamiltonian.
Recalling (\ref{i5}) we have to check that the identity
         \begin{equation}
         {\hbox{\rm Tr}}_{{\cal H}^{(1)}}
         \biggl[
         {{\varrho}^{(1)}(t)
         \left(
         {\mbox{\sf Q}+ {\mbox{{\sf Q}}}^{\scriptscriptstyle \dagger}}
         \right)
         }
         \biggr]
                  =
         -
         {\hbox{\rm Tr}}_{{\cal H}^{(1)}}
         \biggl[
         {
         {\varrho}^{(1)}(t)
         \sum^{}_{{\xi,\lambda }}
        \pi_{\xi}
         {{\mbox{\sf L}}_{\lambda\xi}^{\scriptscriptstyle \dagger}}
         {{\mbox{\sf L}}_{\lambda\xi}}
         }
         \biggr]
         \label{bb29}
         \end{equation}
holds within the approximations so far introduced.
Then we can replace the second term in the l.h.s. of (\ref{2b12}) by
$         {1\over 2\hbar}
        \left \{
        {\sum^{}_{{\xi,\lambda }}
        \pi_{\xi}
         {{\mbox{\sf L}}_{\lambda\xi}^{\scriptscriptstyle \dagger}}
         {{\mbox{\sf L}}_{\lambda\xi}}
         ,{\varrho}^{(1)}(t)}\right \}
$.
Equation~(\ref{bb29}) may be rewritten as
         \begin{equation}
         \sum_{kf}{\varrho}^{(1)}_{fk}(t)
         \left(
         {\mbox{{\sf Q}}+ {\mbox{{\sf Q}}}^{\scriptscriptstyle \dagger}}
         \right)
         _{kf}
         =
         -
         \sum^{}_{{\xi,\lambda \atop g,k,f }}
         {\varrho}^{(1)}_{fk}(t)
         \pi_{\xi}
         {\bigl( {{\mbox{\sf L}}_{\lambda\xi}^{\scriptscriptstyle \dagger}}
         \bigr)}_{kg}
         {\bigl( {{\mbox{\sf L}}_{\lambda\xi}} \bigr)}_{gf}.  \label{bb29a}
         \end{equation}
The part of the
l.h.s. of (\ref{bb29a}) not containing the statistical
operator is equal to
        \begin{equation}
        {\hbox{\rm Tr}}_{{{{\cal H}_{\scriptscriptstyle
        F}}}}
        \left \{
        \left[{
        \left(
        {\cal T}
        \left({-{i \over \hbar} {E_k} +\eta}\right)   {a_{k}}
        \right)
        {a^{\scriptscriptstyle \dagger}_{f}}
        +
        {a_{k}}
        \left(
        {{\cal T}
        \left({{i \over \hbar} {E_f} +\eta}\right)
        {a^{\scriptscriptstyle \dagger}_{f}}
        }
              \right)
        }\right]
        {{\varrho}^{\rm m}(t)}
         \right \}
         .  \label{2b13}
        \end{equation}
The r.h.s. demands a more complex calculation
        {\openup5pt
        \begin{eqnarray*}
        &&
                -
        {1 \over \hbar}
        \sum_{{\xi,\lambda'' \atop g }}
        \pi_{\xi}
         {\bigl( {{\mbox{\sf L}}_{\lambda''\xi}^{\scriptscriptstyle \dagger}}
         \bigr)}_{kg}
         {\bigl( {{\mbox{\sf L}}_{\lambda''\xi}} \bigr)}_{gf}
        =
        \\
        &&
        =-2\eta
	\! \!
        \sum^{}_{{\lambda,\lambda',\lambda''  \atop g}}
        \left \{
        \langle
        {\lambda''}
               \vert
        {
        \left(
        {\cal T}
        \left({-{i \over \hbar} {E_g} +\eta}\right)
           {a_{{g}}}
        \right)
        {a^{\scriptscriptstyle \dagger}_{f}}
        }
                    \vert
        {\lambda}
                         \rangle
        {{\varrho}^{\rm m}_{\lambda\lambda'}(t)}
        \langle
        {\lambda'}
               \vert
        {
        {a_{k}}
        \left(
        {{\cal T}
        \left({{i \over \hbar} {E_g} +\eta}\right)
        {a^{\scriptscriptstyle \dagger}_{g}} }
              \right)
        }
                    \vert
        {\lambda''}
                         \rangle
        \right \}
        \\
        &&
        \hphantom{=}
        \times
        \left[{
        {
        1
        \over
         -{i \over \hbar} {E_g} -\eta  -{i \over \hbar}
         \left(
         {{E_{{\lambda}''}}-{E_f}-{E_{{\lambda}}}    }
         \right)
        }
        +
        {
        1
        \over
         {i \over \hbar}{E_g} -\eta+{i \over \hbar}
         \left(
         {{E_{{\lambda}''}}-{E_k}-{E_{{\lambda}'}}}
         \right)
        }
        }\right]
        \\
        &&
        \hphantom{=\times}\times
        {
        1
        \over
         -2\eta+{i \over \hbar}
         \left(
         {{E_f}+{E_\lambda}-{E_k}-{{E_{\lambda'}}}}
         \right)
        }
        \simeq
        \end{eqnarray*}
        }
having in mind   to demonstrate (\ref{bb29a})
we now rely on (\ref{223})
        \begin{eqnarray*}
        &&
        \simeq
        \sum^{}_{{\lambda,\lambda',\lambda''  \atop g}}
        \langle
        {\lambda''}
        \vert
        \left[{
        \left(
        {-{i \over \hbar} {E_g} -\eta  -{\cal H}_0}
        \right)
        ^{-1}
        \left(
        {{\cal T}
        \left({-{i \over \hbar} {E_g} +\eta}\right)
           {a_{{g}}} }
        \right)
        }\right]
        \vert
        {\lambda f}
        \rangle
        {{\varrho}^{\rm m}_{\lambda\lambda'}(t)}
        \\
        &&
        \hphantom{\mbox{}+{}}\times
        \langle
        {\lambda' k}
        \vert
        \left(
        {{\cal T}
        \left({{i \over \hbar} {E_g} +\eta}\right)
        {a^{\scriptscriptstyle \dagger}_{g}} }
              \right)
        \vert
        {\lambda''}
        \rangle
        +     {}
        \\
        &&
        {}      +
        \sum^{}_{{\lambda,\lambda',\lambda''  \atop g}}
        \langle
        {\lambda''}
        \vert
        \left(
        {{\cal T}
        \left({-{i \over \hbar} {E_g} +\eta}\right)
           {a_{{g}}} }
        \right)
        \vert
        {\lambda f}
        \rangle
        {{\varrho}^{\rm m}_{\lambda\lambda'}(t)}
        \\
        &&
        \hphantom{\mbox{}+{}}\times
        \langle
        {\lambda' k}
               \vert
        {
        \left[{
        \left(
        {{i \over \hbar} {E_g} -\eta  -{\cal H}_0}
        \right)
        ^{-1}
        \left(
        {{\cal T}
        \left({{i \over \hbar} {E_g} +\eta}\right)
        {a^{\scriptscriptstyle \dagger}_{g}} }
        \right)
        }\right]
        }
                    \vert
        {\lambda''}
                         \rangle
        ,
        \end{eqnarray*}
but using the identity
        \[
        {{
        \left(
        {{ z-\eta - {\cal H}_0}}
        \right)
        }^{-1}}
        {\cal T}(z+\eta)=
        \left({1+2\eta
        {{
        \left({{ z-\eta - {\cal H}_0}}\right)
        }^{-1}}
        }\right)
        \left({
        {{
        \left({{ z+\eta - {\cal H}}}\right)
        }^{-1}}
        {\cal V}}\right)
        \]
we get to zero order in $\eta$,
        \[
        \left(
        { -{i \over \hbar} {E_g} -\eta  -{\cal H}_0}
        \right)
        ^{-1}
        \left({{\cal T}
        \left({-{i \over \hbar} {E_g} +\eta}\right)
        {a_{{g}}} }\right)
        = {a_{{g}}},
        \]
and similarly
        \[
        \left({+{i \over \hbar} {E_g} -\eta  -{\cal H}_0}\right)
        ^{-1}
        \left({{\cal T}
        \left({+{i \over \hbar} {E_g} +\eta}\right)
        {a^{\scriptscriptstyle \dagger}_{g}} }\right)
        = {a^{\scriptscriptstyle \dagger}_{g}},
        \]
thus obtaining
        \begin{eqnarray*}
        &&
        -
        {1 \over \hbar}
        \sum^{}_{{\xi,\lambda \atop g }}
        \pi_{\xi}
        {\bigl( {{\mbox{\sf L}}_{\lambda\xi}^{\scriptscriptstyle \dagger}}
        \bigr)}_{kg}
        {\bigl( {{\mbox{\sf L}}_{\lambda\xi}} \bigr)}_{gf}
        =
        \\
        &&=
        {\hbox{\rm Tr}}_{{{{\cal H}_{\scriptscriptstyle
        F}}}}
        \left \{
        {
        \left[{
        \left({{\cal T}
        \left({-{i \over \hbar} {E_k} +\eta}\right)
           {a_{k}}
        }\right)
        {a^{\scriptscriptstyle \dagger}_{f}}
        +
        {a_{k}}
        \left({{\cal T}
        \left({{i \over \hbar} {E_f} +\eta}\right)
        {a^{\scriptscriptstyle \dagger}_{f}}
        }\right)
        }\right]
        {{\varrho}^{\rm m}(t)}
        }\right \},
        \end{eqnarray*}
that is to say the same expression as in  (\ref{2b13}).
\par                            
\section{PHYSICAL DISCUSSION AND CONCLUSIVE REMARKS}
\par
\setcounter{equation}{0}
\vskip 10pt
To elucidate how an equation of the form (\ref{2b12}) or
equivalently (\ref{2b10})
may be well suited to describe an interplay between a ``purely
optical'' (that is wavelike) dynamics and an interaction with a
measuring character let us introduce the reversible mappings
$
{\cal A}_{t'' t'}={U}_{t'' t'} \cdot
{U}^{\scriptscriptstyle \dagger}_{t'' t'}
$,
  where
        \begin{equation}
        {U}_{t'' t'}
        = T
        \left({e^{-{i \over \hbar} \int_{t'}^{t''} d\tau \,
        \left({\mbox{\sf H}_0 (\tau)+i\mbox{{\sf Q}}(\tau)}\right)
        }}\right)
        ,
         \label{31}
        \end{equation}
corresponding to a coherent contractive evolution of the microsystem during
the time interval $[t',t'']$, and
the
completely
positive mappings
        \begin{equation}
        {\cal L}_{\lambda \xi}=
        {\mbox{\sf L}}_{\lambda\xi}(t) \cdot
        {{\mbox{\sf L}}_{\lambda\xi}^{\scriptscriptstyle \dagger}}(t)
        \pi_{\xi(t)}
        ,  \label{32}
        \end{equation}
whose measuring character may be inferred from the discussion
following (\ref{misura}).
The structure of the operators   ${\mbox{\sf L}}_{\lambda\xi}$
[see (\ref{2b11})]
further shows that these mappings may be linked with a transition
inside the macrosystem
 specified by the pair of indexes
$\xi,\lambda$,
as a result of scattering with the microsystem.
Under very particular conditions,
strongly  enhancing the measuring character of the interaction
(as would be the case for a detector), these transitions could be
macroscopic detectable, thus leading to a localization of the
particle. To indicate such interactions we will therefore use the
word ``event''.
\par
The solution of  (\ref{2b12}) can be
written as:
        \begin{eqnarray}
        \label{33}
        {\varrho}_t
        &=&
        {\cal A}_{t t_0}{\varrho}_{t_0} +
        \sum_{\lambda_1 \xi_1}
        \int_{t_0}^{t} dt_1 \, {\cal A}_{t t_1} {\cal L}_{\lambda_1
        \xi_1}(t_1)
        {\cal A}_{t_1 t_0}{\varrho}_{t_0} +{}
        \nonumber \\
        {}&&
        +
        \sum_{\lambda_1 \xi_1 \atop \lambda_2 \xi_2}
        \int_{t_0}^{t} dt_2 \,
        \int_{t_0}^{t_2} dt_1 \,
        {\cal A}_{t t_2} {\cal L}_{\lambda_2 \xi_2}(t_2)
        {\cal A}_{t_2 t_1}
        {\cal L}_{\lambda_1 \xi_1}(t_1)
        {\cal A}_{t_1 t_0}{\varrho}_{t_0} +
        \ldots
        \end{eqnarray}
which can be interpreted as a sum over subcollections
corresponding to the realization of no event, one event, two
events and so on.
To see this let
us perform some measurement on the microsystem at time $t$,
associated with an eigenstate $u_\alpha$ of some observable
$A$.
Then by
(\ref{32}) and (\ref{33}) the probability $p_{\alpha}(t)$ of the result
${\alpha}$ for this observable at time $t$ has the following
structure:
        \begin{eqnarray}
        \label{34}
        \!\!\!\!\!\!
        p_{\alpha}(t)
        & = &
        \langle
        {u_{\alpha}}
        \vert
        {{\cal A}_{t t_0}{\varrho}_{t_0}}
        \vert
        {u_{\alpha}}
        \rangle
         +
        \sum_{\lambda_1 \xi_1}
        \int_{t_0}^{t} dt_1 \,
        \langle
        {u_{\alpha}}
        \vert
        {{\cal A}_{t t_1} {\cal L}_{\lambda_1 \xi_1}(t_1)
        {\cal A}_{t_1 t_0}{\varrho}_{t_0}}
        \vert
        {u_{\alpha}}
        \rangle
         +{}
        \nonumber \\
        {}&&
        +
        \sum_{\lambda_1 \xi_1 \atop \lambda_2 \xi_2}
        \int_{t_0}^{t} dt_2 \,
        \int_{t_0}^{t_2} dt_1 \,
        \langle
        {u_{\alpha}}
               \vert
        {{\cal A}_{t t_2} {\cal L}_{\lambda_2 \xi_2}(t_2)
        {\cal A}_{t_2 t_1}
        {\cal L}_{\lambda_1 \xi_1}(t_1)
        {\cal A}_{t_1 t_0}{\varrho}_{t_0}}
                    \vert
        {u_{\alpha}}
                         \rangle
         +
        \ldots
        \end{eqnarray}
Let us assume for simplicity that the initial preparation
${\varrho}_{t_0}$ is a pure state
${\varrho}_{t_0}=
{\vert  {\psi_{t_0}}   \rangle}
{\langle \psi_{t_0} \vert}
$, then the first term
in the l.h.s. of  (\ref{34}) has by (\ref{31}) the form:
        \begin{equation}
        \langle
        {u_{\alpha}}
        \vert
        {{\cal A}_{t t_0}{\varrho}_{t_0}}
        \vert
        {u_{\alpha}}
        \rangle
        =
        {  \left \vert
        {\langle {u_{\alpha}}   \vert {\psi(t)}  \rangle}
        \right \vert
        }^2
        ,
        \quad
        \psi(t)=
        T
        \left({e^{-{i \over \hbar} \int_{t_0}^{t} d\tau \,
        \left({\mbox{\sf H}_0(\tau)+i\mbox{{\sf Q}}(\tau)}\right)
        }}\right)
         \psi_{t_0} ,
         \label{35}
        \end{equation}
and it gives the probability of measuring  $A={\alpha}$
at time $t$ when no event is produced in between the preparation
of the state $\psi_{t_0}$ at time $t_0$ and the measurement of
$A$ at time $t$; the trace of the first subcollection $p_t^0 =
{\hbox{\rm Tr}}_{{\cal H}^{(1)}}
{{\cal A}_{t t_0}{\varrho}_{t_0}} =
{\left\| \psi(t) \right\|}^2$ gives
the probability that no event happens in the time interval
$[t_0,t]$; then apart from the fact that $p_t^0 \leq 1$ ($p_t^0 $
is a non-increasing function) the usual statistical
interpretation of the wave-function is recovered. The integrand
of the second term
$
        \langle
        {u_{\alpha}}
        \vert
        {{\cal A}_{t t_1} {\cal L}_{\lambda_1 \xi_1}(t_1)
        {\cal A}_{t_1 t_0}{\varrho}_{t_0}}
        \vert
        {u_{\alpha}}
        \rangle
$
can be interpreted as the probability   of detecting $A={\alpha}$
at time $t$, when 
the transition $\lambda_1
\xi_1$ happens in the time interval $[t', t' + dt']$, while no
transition $\lambda \xi$ happens in the time intervals $[t_0, t']$,
$[t' + dt',t]$; in other words the expression
        \[
        \int_{t_0}^{t} dt_1 \,
        \langle
        {u_{\alpha}}
        \vert
        {{\cal A}_{t t_1} {\cal L}_{\lambda_1 \xi_1}(t_1)
        {\cal A}_{t_1 t_0}{\varrho}_{t_0}}
        \vert
        {u_{\alpha}}
        \rangle
        \]
gives the probability of
$A={\alpha}$
at time $t$ when one and only one event linked to the transition
$\lambda_1
\xi_1$ happens in the time interval $[t_0, t]$, while
        \[
        p^1_{t}=
        {\hbox{\rm Tr}}_{{{\cal H}^{(1)}}}
        \left({\int_{t_0}^{t} dt_1 \, {\cal A}_{t t_1}
        {\cal L}_{\lambda_1 \xi_1}(t_1)
        {\cal A}_{t_1 t_0}{\varrho}_{t_0} }\right)
        \]
is just the probability for this sole event in the time  interval
$[t_0, t]$.
While the first term in the l.h.s. of (\ref{33})  is a pure state,
provided ${\varrho}_{t_0}$ is, the second one, due to different
transition times, is a mixture.
The other terms of (\ref{33}) provide the almost obvious
generalization describing  repeated production of events
$\lambda\xi$.
\par
If the macrosystem is an interferometer, the
role of the first term is enhanced by the experimental situation,
nevertheless if one can monitor the path followed by the
microsystem inside the interferometer, then the other terms also
become relevant. If at the output of the interferometer an
interference pattern is observed, some disturbance by an
incoherent background due to these terms is unavoidable.
Obviously such disturbance can be made negligible if the
experimental set-up is such as to ``automatically'' select only
coherent contributions. This is the case if the disturbance
originates in scattering and the acceptance along the whole path
is small enough as in
neutron-interferometry, however, forward scattering
cannot be eliminated, so, even simply relying on the present general
theoretical framework, one should expect that the first term of
(\ref{34}) cannot account for the whole experimental evidence, and
this could explain some difficulties that have been reported in
the interpretation of neutron interference experiments,
without resorting to a reformulation of quantum
mechanics, as proposed by Namiki and Pascazio (1993).
A more precise insight into the structure of the operators $\mbox{{\sf Q}}$
and ${\mbox{\sf L}}$ can be obtained introducing the field operator
        \[
        \psi({\mbox{\bf x}},\omega)=\sum_f {a_{{f}}} u_f({\mbox{\bf x}},\omega)
        ,
        \quad
        {a_{{f}}} = \sum_{\omega}
        {\int d^3 \! x \,}
        u_f^*({\mbox{\bf x}},\omega)        \psi({\mbox{\bf x}},\omega)
        \]
and writing instead of (\ref{213}):
        \[
        \left({{\cal T}(z)\psi}\right)
        ({\mbox{\bf x}},\omega)
        =
        \sum_{\omega'}
        {\int d^3 \! x' \,}
        {\rm T}
        \left({{\mbox{\bf x}},\omega,{\mbox{\bf x}}',\omega',z}\right)
         \psi({\mbox{\bf x}}',\omega')
        .
        \]
Then  (\ref{212a}) becomes:
        \[
  	T{}_{l}^{k}
        \left(
        z
        \right)
        =
        \sum_{\omega,\omega'}
        {\int d^3 \! x \,}
        { d^3 \! x' \,}
        u_k^*({\mbox{\bf x}},\omega)
        {\rm T}
        \left({{\mbox{\bf x}},{\mbox{\bf x}}',\omega,\omega',z}\right)
        u_l({\mbox{\bf x}}',\omega')   ,
        \]
and assuming translation invariance
	\begin{eqnarray}
        \label{38}
                T{}_{l}^{k}
        \left(
        z
        \right)
        &=&
        \sum_{\omega,\omega'}
        {\int d^3 \! x \,}
        { d^3 \! x' \,}
        u_k^*({\mbox{\bf x}},\omega)
        {\rm T}
        \left({{\mbox{\bf x}}-{\mbox{\bf x}}',\omega,\omega',z}\right)
        u_l({\mbox{\bf x}}',\omega')
         =
        {\int d^3 \! X \,}
        T{}_{l}^{k}
        \left(
        {{\mbox{\bf X}},z}
        \right)
        ,
        \nonumber \\
        T{}_{l}^{k}
        \left(
        {{\mbox{\bf X}},z}
        \right)
        &=&
        \sum_{\omega,\omega'}
        {\int d^3 \! r \,}
        u_k^*({\mbox{\bf X}}+ {\mbox{\bf r}\over 2},\omega)
        {\rm T}
        \left({\mbox{\bf r},\omega,\omega',z}\right)
        u_l({\mbox{\bf X}}- {\mbox{\bf r}\over2},\omega')
         .
        \end{eqnarray}    

In correspondence with the representation (\ref{38}) of
$
        T{}_{l}^{k}
        \left(
        z
        \right)
$ one has a similar representation for
$
{\bigl( {{\mbox{\sf L}}_{\lambda\xi}} \bigr)}_{kf}
$:
        \begin{equation}
        {\bigl( {{\mbox{\sf L}}_{\lambda\xi}} \bigr)}_{kf}
        =
        {\int d^3 \! X \,}
        {\bigl[ {{\mbox{\sf L}}_{\lambda\xi}}({\mbox{\bf X}}) \bigr]}_{kf},
        \label{38bis}
        \end{equation}
simply obtained substituting (\ref{38}) inside
(\ref{2b11}).
\par
The set of variables $N_{\lambda\xi}(\tau)$, $\tau\geq t_0$,
$N_{\lambda\xi}(\tau)$ being the number of transitions $\lambda\xi$
up to
time $\tau$, define a multicomponent classical stochastic process
for which   probability distributions and description of
statistical subcollections at times $\tau$, conditioned
by the values $N_{\lambda\xi}(\tau)$, can be given.
This is a straightforward generalization of the typical
``counting process'' considered by Srinivas and
Davies
(1981); e.g. the probability that in a time interval
$[\tau_1,\tau_2]$ there are $N$ events related to transitions
$\lambda_1\xi_1,\lambda_2\xi_2,\ldots,\lambda_N\xi_N$($\vec\lambda
\vec\xi$),
belonging respectively  to  certain subsets
$
\sigma_1 \in \Gamma_{t_1},
\sigma_2 \in \Gamma_{t_2},
\ldots
\sigma_N \in \Gamma_{t_N}
$
($\lambda$ and $\xi(t)$ belong respectively to the spectra
$\Lambda$ of ${H}_{\rm m} $ and $\Xi(t)$ of ${{\varrho}^{\rm m}(t)}$,
which are
practically a continuum, and $\Gamma_t$ is a $\sigma$-algebra on
$\Lambda \times \Xi(t)$),
when no event happens
before $\tau_1$, is given by:
        \[
        P_{\tau_1,\tau_2}(N,\vec\sigma) =
        {\hbox{\rm Tr}}
        \left({
        {\cal F}_{\tau_1,\tau_2}(N,\vec{\sigma})
        {{\cal A}_{t t_0}{\varrho}_{t_0}}
        }\right)
        \]
where ${\cal F}_{\tau_1,\tau_2}(N,\vec{\sigma})$ is an operation,
i.e. a
contractive positive mapping on ${\cal T}({{\cal H}^{(1)}})$:
        \begin{eqnarray*}
        {\cal F}_{\tau_1,\tau_2}(N,\vec{\sigma})
        &=&
        \\
        &=&
        \sum_{
        \left({\vec\lambda\vec\xi}\right)
        \in\vec{\sigma}}
        \int_{\tau_1}^{\tau_2}
        dt_N \ldots
        \int_{\tau_1}^{t_2}
        dt_1
        {\cal A}_{\tau_2 t_N} {\cal L}_{\lambda_N \xi_N}(t_N)
        {\cal A}_{t_N t_{N-1}}
        \ldots
        {\cal L}_{\lambda_1 \xi_1}(t_1)
        {\cal A}_{t_1 \tau_1}.
        \end{eqnarray*}
This flow of transitions  accompanying in the medium the propagation
of
the microsystem could prime a measurement inside some suitable
measuring device, then $P_{\tau_1,\tau_2}(N,\vec{\sigma})$ would be
the
probability for this device to be affected by the microsystem. In
fact writing ${F}(\vec{\sigma})={{\cal F}'}_{\tau_1,\tau_2}(N,
\vec{\sigma}){I}$,
with ${\cal F}'$ the adjoint mapping on ${\cal B}({{{\cal
H}}^{(1)}})$, (the set of bounded operators on ${{{\cal
H}}^{(1)}}$) one has:
        \begin{equation}
        P_{\tau_1,\tau_2}(N,\vec{\sigma})
        =
        {\hbox{\rm Tr}}_{{\cal H}^{(1)}}
        \left({{F}(\vec{\sigma}){\cal A}_{t
        t_0}{\varrho}_{t_0}}\right)
        ,
         \label{310}
        \end{equation}
${F}(\vec{\sigma})$ being a positive operator,
${F}(\vec{\sigma})\leq 1$. Equation~(\ref{310}) is the typical
probability
rule of modern quantum mechanics in which the notion of an
``effect valued measure'' ${F}(\vec{\sigma})$ on some
$\sigma$-algebra of subsets generalizes the customary concept of
a projection valued measure, or equivalently of a self-adjoint
operator, associated to an observable; these observables
present an idealisation that is very useful to understand the
basic structure of quantum mechanics, but is too strong for
representing real measuring
devices
(Ludwig, 1983; Kraus, 1983; Holevo, 1982; Davies, 1976).
A similar
situation
is met if one considers the statistical operator
        \[
        {\varrho}_{\tau_2}
        =
        {
        {\cal F}_{\tau_1,\tau_2}(N,\vec{\sigma})
        {\cal A}_{t t_0}{\varrho}_{t_0}
        \over
        P_{\tau_1,\tau_2}(N,\vec\sigma)
        } ,
        \]
which represents the repreparation at time $\tau_2$ of the
statistical collection ${\varrho}_{t_0}$ under the condition that the
aforementioned effect happens in the time interval
$[\tau_1,\tau_2]$.
Taking  (\ref{32}) into account ${\varrho}_{\tau_2}$
is seen to bear an analogy with the highly idealized von-Neumann
state reduction rule
        \[
        {\varrho}_{\tau_2}^{\scriptscriptstyle(+)}
        =
        {
        {P}{\varrho}_{\tau_2}^{\scriptscriptstyle(-)}
        {P}
        \over
        {\hbox{\rm Tr}}
        \left({{P}{\varrho}_{\tau_2}^{\scriptscriptstyle(-)}}\right)
        }
        \]
for the statistical operator
${\varrho}_{\tau_2}^{\scriptscriptstyle(-)}$, when it is reprepared at
time $\tau_2$ taking a measurement into account, associated with
the projection operator ${P}$.
\par
Actually by (\ref{33}) a decomposition of ${\varrho}_t$ is given into
subcollections related to all possible detection patterns of
events primed by the elementary transitions $\lambda\xi$;
mathematically this means that a decomposition of the evolution
mapping ${\rm T}
\left({\exp \int_{t_0}^{t} \, d t'  \, {\cal L}(t')}\right)
$
has been given on the space of the jump processes
$N_{\lambda\xi}(\tau)$.
In different physical contexts, e.g. optical heterodyne detection,
more 
general decompositions of an evolution
mapping can be given, as it has been shown in the aforementioned
theory of continuous
measurement: then
the variables involved are not only
$N_{\lambda\xi}(\tau)$, but also the values of continuously measured
variables related to the system.
\par
{
\parindent = 0 pt
\vskip 20pt
{\LARGE References}
\vskip 20pt
\par
{Comi, M., Lanz, L., Lugiato, L.~A., and
Ramella, G.}
(1975).
{\it Journal of Mathematical Physics},
{\bf 16},
{910}.

{Chapman, M.~S., {et al.}} (1995).
{\it Physical Review Letters},
{\bf 75},
{3783}.

{Davies, E.~B.}
(1976).
{\it Quantum Theory of Open
Systems}, {Academic Press}, {London}.

Holevo, A.~S.
(1982).
{\it Probabilistic and Statistical Aspects of
Quantum Theory}, {North Holland}, {Amsterdam}{}.

Kraus, K.
(1983).
States, Effects and Operations, in {Lecture Notes in
Physics}, Volume 190, {Springer}, {Berlin}.

{Lanz, L.}
(1994).
{\it International Journal of Theoretical Physics},
{\bf 33},
{19}.

Lanz, L., and Melsheimer, O.
(1993). 
Quantum Mechanics and Trajectories,
in
{\it Symposium On the Foundations of Modern Physics},
Busch, P., Lahti, P.~J., and Mittelstaedt, P., eds., World
Scientific, Singapore, p.233-241.

{Lindblad, G.}
(1976).
{\it Communications in Mathematical Physics},
{\bf 48},
{119}.

{Ludwig, G.}
(1976).
{\it Einf\"uhrung in die Grundlagen der
Theoretischen Physik}, {Vieweg}, {Braunschweig}{, Band 3, Quantentheorie}.

{Ludwig, G.}
(1983).
{\it Foundations of Quantum
Mechanics}, {Springer}, {Berlin}.

Mittelstaedt, P., Prieur, A., and Schieder, R.
(1987).
{\it Foundations of Physics},
{\bf 17},
{891}.

{Namiki, M., and Pascazio, S.}
(1991).
{\it Physical Review A},
{\bf 44},
{39}.

{Namiki, M., and Pascazio, S.}
(1993).
{\it Physics Reports},
{\bf 232},
{301}.

Namiki, M., Pascazio S., and Rauch, H.
(1993).
{\it Physics Letters A},
{\bf 173},
{87}.

Rauch, H. (1990). In {\it Proc.~3nd Int.~Symp. on Foundations
of Quantum Me\-cha\-nics}, S.~Ko\-ba\-ya\-shi {\it et al.}, eds.,
Physical Society, Tokyo, p.3

Rauch, H. (1995). In {\it Advances in Quantum Phenomena}, 
E.~G.~Beltrametti and J.-M.~L\'evy-Leblond, eds., NATO ASI series,
Vol.~B347, Plenum Press, New York, p.113.

Rauch, H., Summhammer, J., Zawisky, M., and Jericha, E.
(1990).
{\it Physical Review A},
{\bf 42},
{3726}.

Sears, V.~F.
(1989).
{\it Neutron Optics},
Oxford University Press, Oxford.

{Srinivas, M.~D., and Davies, E.~B.}
(1981).
{\it Optica
Acta},
{\bf 28},
{981}.

{Taylor, J.~R.}
(1972).
{\it Scattering Theory},
{John Wiley and Sons}, {New York}.

{Thomson, M.~J.}
(1993).
{\it Physics Letters A},
{\bf 179},
{239}.

{Vigu\'e, J.}
(1995).
{\it Physical Review A},
{\bf 52},
{3973}.
}

\end{document}